\documentclass[slac_one]{revtex4}
\usepackage{graphicx}
\usepackage{fancyhdr}
\newcommand{\Zo}{\mathrm{Z}}

\begin{document}

\title{{\small{2005 International Linear Collider Workshop - Stanford,
U.S.A.}} \\ 
\vspace{12pt}

Z$'$ Signals from Kaluza-Klein Dark Matter} 

%

\author{S. Riemann}
\affiliation{DESY, Platanenalle 6, D-15738 Zeuthen, Germany}

\begin{abstract}
In a minimal model of universal extra dimensions the lightest Kalzua-Klein particle is neutral and stable because of Kaluza-Klein parity conservation and it is a dark matter candidate.
The corresponding level 2 mode of the lightest Kaluza-Klein particle can be detected  via KK number violating decays to Standard Model particles. The expected sensitivity to second level Kaluza-Klein modes of neutral gauge bosons at the ILC is studied.

\end{abstract}

\maketitle
 
\thispagestyle{fancy}

\section{INTRODUCTION} 
The ideas of new physics models with extra dimensions offer a rich phenomenolgy and predict signals that could be measured in future collider experiments.
An interesting scenario of universal extra dimensions (UED) is suggested by \cite{ref:acd}: 
All Standard Model fields are placed in the bulk and propagate in extra dimensions of size $1/R\sim 1~$TeV. To all Standard Model particles exist Kaluza-Klein (KK) partners  with identical spins and couplings and highly degenerate mass spectra;
\begin{equation}\label{eq:mass-spectra}
M^2_n=\frac{n^2}{R^2}+M_0^2
\end{equation}
where $R$ is the (unknown) compactification radius.
The minimal UED (MUED) model is defined in five dimensions, the extra dimension is compactified on a $S^1/Z_2$ orbifold.
Due to momentum conservation 
the KK number is conserved in all processes and at tree level pairs of Standard fermions are not allowed to interact directly with higher modes of gauge boson KK towers.  

Interactions at localized fixed points of the orbifold violate the 5D Lorentz invariance and 
lead to a conservation of KK parity, $(-1)^n$. This implies that the lightest KK particle (LKP) is stable and level 1 KK gauge bosons are not allowed to decay directly to two Standard model particles. 
The KK number violating decay of a level 2 gauge boson into two  Standard Model particles is possible.
The LKP is a dark matter candidate.  

Precision electroweak measurements constrain the  bounds on $R$, $1/R \geq 250$~GeV \cite{ref:acd}. Cosmological bounds suggest $1/R < {\cal{O}}(1)$~TeV for $\Omega \sim 0.3$ \cite{ref:cosm-bound}. With these limits there is a good chance to observe KK particles with future collider experiments.

The mass degeneracy (Equ. \ref{eq:mass-spectra}) is only valid at tree level and is lifted by radiative corrections \cite{ref:matchev-radcor,ref:matchev-bos-susy,ref:rizzo-ued}.
The bulk corrections are finite, $\Delta M_n \sim 1/(16 \pi^4 R^2)$, the correction from the boundary terms are logarithmically divergent, $\Delta M_n \sim M_n/16\pi^2 \ln (\Lambda^2/\mu^2)$ with the cut-off scale $\Lambda$ and the renormalization scale $\mu\approx M_n$.
%
The KK number violating couplings are related to the 
mass corrections from the boundary terms and are much smaller than the corresponding Standard Model couplings (loop suppressed).

The first level KK modes can be obtained at the ILC by the processes $e^+ e^- \rightarrow e^+_{(1)} e^-_{(1)}$ \cite{ref:level1-ee} or $e^+ e^- \rightarrow \mu^+_{(1)} \mu^-_{(1)}$ \cite{ref:level1-mm}. The energy splitting between $\gamma_{(1)}$ and the level 1 leptons allows the decays  to the final states $e^+e^-$ or $\mu^+\mu^-$  and 2 $\gamma_{(1)}$'s carrying away missing energy. 


\section{KK LEVEL 2 GAUGE   BOSON EXCHANGE IN $e^+e^- \rightarrow f\bar{f}$}

KK gauge bosons can be produced pairwise through KK number preserving interactions. 
The single production of KK gauge bosons is restricted by KK parity conservation. As a consequence   
the exchange of level 1 KK gauge bosons is not allowed in fermion pair production but 
level 2 photons or Z bosons can be exchanged through KK number violating couplings to Standard Model quarks and leptons. 
An ILC running at the resonance of a second level photon or Z boson would be perfect for measuring the spins and the mass splittings between the states. But most likely the required c.m.s. energy is too high and   only an indirect observation could be possible.
The search for level 2 gauge bosons is similar to the usual search for Z$'$ bosons - assuming very small
couplings to fermions.
Below the resonances the modification of the hadronic and leptonic cross section,
\begin{equation}\label{equ:xs}
\frac{d \sigma}{d \cos \theta} \sim \sum \left( A_{ij}^{SM}+\frac{Q_{\gamma_{(2})}^e Q_{\gamma_{(2)}}^f}{s-M^2_{\gamma_{(2)}}+iM_{\gamma_{(2)}}\Gamma_{\gamma_{(2)}}}+\frac{g_i^{\Zo_{(2)},e} g_j^{\Zo_{(2)},f}}{s-M^2_{\Zo_{(2)}}+iM_{\Zo_{(2)}}\Gamma_{\Zo_{(2)}}}\right)^2 \rho(\cos \theta) d\cos \theta
\end{equation}
can be measured and used to determine the compactification radius, $1/R$. 

\subsection{Bounds on Z$_2$ and $\gamma_2$ }

To determine the sensitivity for an indirect detection of level 2 gauge bosons with fermion pair production an integrated luminosity of 1~ab$^{-1}$ with $\delta L_{int}=$0.1\% is assumed. 
\begin{figure*}[h]
\centering
\includegraphics[width=135mm,height=135mm]{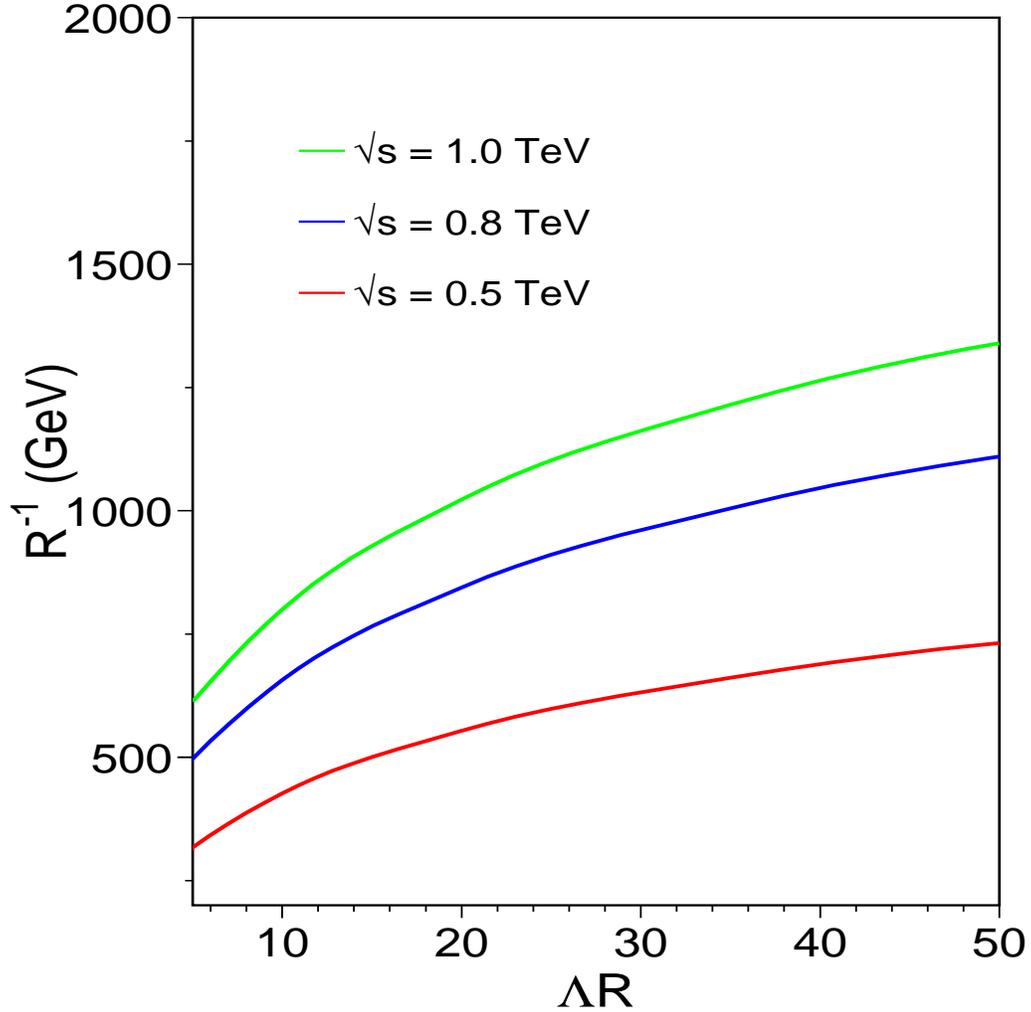}
\caption{Limits at 95\% C.L. on 1/R derived from the combined measurements of leptonic and hadronic final states. Deviations from the SM expectations can be observed up to the given values. } \label{r_lep+had}
\end{figure*}
Uncertainties due to  the identification of leptonic and hadronic final states are assumed to result in  syatematic errors of 0.1\%. A  polarisation of 80\% for the electron and 60\% for the positron beam and an error of 0.1\% for the polarisation measurement is considered. 50\% of the luminosity are spent to each sign combination, (+,-) and  (-,+), of beam polarisation.

\begin{figure*}[h]
\centering
\includegraphics[width=135mm,height=135mm]{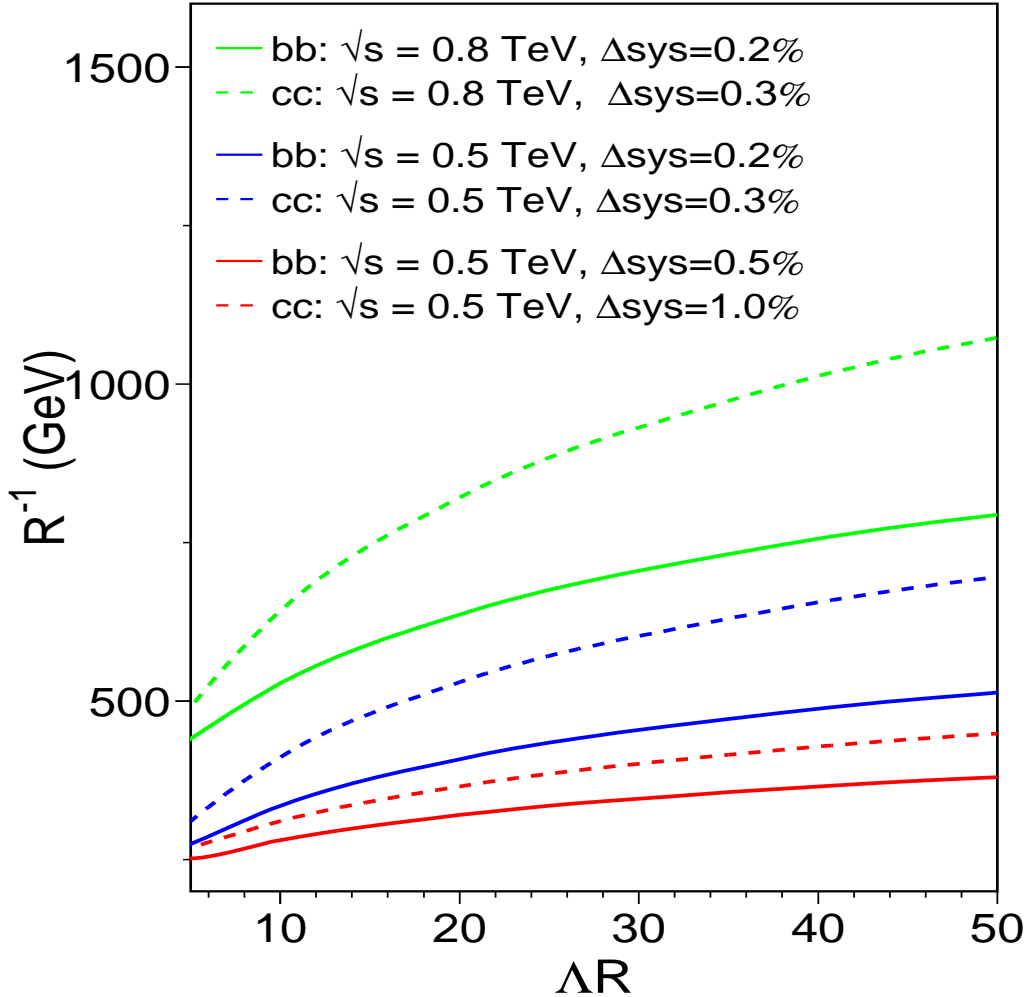}
\caption{Limits at 95\% C.L. on 1/R derived from measurements of $b\bar{b}$ or $c \bar{c}$ final states assuming different systematic errors. Deviations from the SM expectations can be observed up to the given values.} \label{r_bb+cc}
\end{figure*}

A combined fit to the leptonic and hadronic cross sections and asymmetries results in sensitivities on $1/R$ at the 95\% C.L. as shown in Figure \ref{r_lep+had}. For $\Lambda R=20$ an indirect signal can be obtained for $1/R \approx \sqrt{s}$, corresponding to $\gamma_{(2)}, ~\Zo_{(2)}< 2 \sqrt{s}$. This is equivalent to $\gamma_{(1)} < \sqrt{s}$ and exceeds the kinematic limit of direct searches for level 1 KK particles by a factor 2.  
Considering heavy quark final states seperately, the sensitivity given in Figure \ref{r_bb+cc} can be reached depending on the sytematic error of the measurements. For $\Lambda R = 20$ a deviation from the Standard Model predictions can be obtained if $\gamma_{(2)}, ~\Zo_{(2)}<  \sqrt{s}$.

\subsection{Distingiush UED KK bosons from other new physics}
The indirect search for level 2 gauge bosons is similar to a Z$'$ search.
A sequential Z$'$ with couplings as the Standard Z boson and a mass smaller than 16 TeV  would cause observable deviations from the Standard Model predictions at the ILC operating at 1 TeV. The distinction between Z$'$ and level 2 gauge bosons is difficult and also model dependent. In contrast to the usual Z$'$ exchange in a the MUED model both, $\gamma_{(2)}$ and Z$_{(2)}$ are exchanged. Exploiting the modification of the angular distributions or, correspondingly, the asymmetries, it will be possible to distinguish KK bosons from usual Z$'$ bosons. More studies are needed in order to quantify the resolution power. 

\section{SUMMARY}
It has been shown for the model of MUED that precise measurements at the ILC allow the detection of higher KK gauge boson modes also below their direct production threshold. In particular, assuming $R\Lambda=20$ a sensitivity to deviations from Standard model measurements is expected up to a compactification radius $1/R \approx \sqrt{s}$. This corresponds to a  LKP $\gamma_{(1)}\sim \sqrt{s}$. The search for these new particles is similar to a Z$'$ search. Further studies are needed to examine the possibilities to distingiush between the new physics models. The observation of KK particles at the LHC is non-trivial so that measurements at the ILC will complete the LHC searches.

\subsection{Acknowledgments} \label{Ack}
I would like to  thank  K. M\"onig for fruitful discussions.

\end{document}